\documentstyle[preprint,cite,aps]{revtex}
\newcommand{\be}{\begin{equation}}
\newcommand{\ee}{\end{equation}}
\newcommand{\bea}{\begin{eqnarray}}
\newcommand{\eea}{\end{eqnarray}}
\newcommand{\ba}{\begin{array}}
\newcommand{\ea}{\end{array}}

\newcommand{\bb}{\bibitem}

\begin{document}
\draft
\tightenlines

\title{\bf Specific heat amplitude ratio for anisotropic Lifshitz
critical behaviors}
\author{Marcelo M. Leite\footnote{e-mail:leite@fis.ita.br}}
\address{{\it Departamento de F\'\i sica, Instituto
Tecnol\'ogico de Aeron\'autica, Centro T\'ecnico Aeroespacial,
12228-900, S\~ao Jos\'e dos Campos, SP, Brazil}}
\maketitle

\vspace{0.2cm}
\begin{abstract}
{\it We determine the specific heat amplitude ratio near a $m$-axial
Lifshitz point and show its universal character. Using a recent
renormalization group picture along with new field-theoretical
$\epsilon_{L}$-expansion techniques, we established this amplitude ratio
at one-loop order. We estimate the numerical value of this amplitude ratio
for $m=1$ and $d=3$. The result is in very good agreement with its
experimental measurement on the magnetic material $MnP$. It is shown that in
the limit $m \rightarrow 0$ it trivially reduces to the Ising-like amplitude
ratio.}
\end{abstract}

\vspace{1cm}

\newpage
\section{Introduction}

Among all properties of systems near a second order phase transition,
amplitude ratios of certain thermodynamic potentials above and below the
critical temperature play a fundamental role together with the critical
exponents: they are examples of universal quantities. All universal amounts
are characterized by a limited number of parameters defining the universality
class and do not depend on the microscopic details of the system under
consideration. One particular type of universality class is associated to
the Lifshitz critical behavior \cite{Ho-Lu-Sh,Ho} where the inclusion of
competing interactions along one or more spatial directions is the main
difference with respect to the ordinary critical behavior.

Recent investigations of the associated critical exponents for the
$m$-axial Lifshitz universality class have been put forward
using numerical Monte Carlo simulations \cite{Pleim} and field-theoretic
approaches \cite{Diehl,leite1,leite2}. Within the
perturbative $\epsilon_{L}$-expansion, there are two proposals in order
to unravel the higher loop structure of this sort of critical behavior.
One of them makes use of a semi-analytic $\epsilon_{L}$-expansion for
the critical exponents, where some loop integrals are evaluated through
numerical integration \cite{Diehl}. Another alternative is the purely
analytical treatment of all loop integrals involved, such that new
renormalization group as well as $\epsilon_{L}$-expansion ideas have been
developed in order to determine those critical indices
\cite{leite1,leite2}. We shall focus our attention in the latter thoughout
this article for convenience.

The Lifshitz multicritical point arises in a variety of real
physical systems, but we shall be concerned here only with its manifestation
in magnetic systems where it was originally discovered \cite{Ho-Lu-Sh}.
The uniaxial case $m=1$ can be explained in terms of an Ising model with
ferromagnetic interactions among nearest neighbors spins as well as
additional antiferromagnetic couplings among the second neighbors along a
single axis, known as ANNNI model \cite{Se}. The competition originates
several modulated phases in addition to the usual paramagnetic
and ferromagnetic phases. It is very simple to analyse the situation near
the uniaxial Lifshitz point, which arises at the confluence of a modulated
and a ferromagnetic phase with the paramagnetic phase. Allowing these
competing interactions along $m$ spatial directions, one obtains the
$m$-axial Lifshitz critical behavior.  The anisotropic
Lifshitz universality classes associated to this sort of critical behavior
depend upon the number of components of the order parameter $N$, the space
dimension $d$ and the number of competing axes $m$ of the system, therefore,
extending the Ising-like universality classes characterized solely
by ($N,d$). An interesting feature of the critical exponents of such systems
is the property of universality class reduction: in the limit
$m \rightarrow 0$ the critical indices reduce trivially to those
corresponding to the usual Ising-like behavior \cite{leite1,leite2,AL1}.

The theoretical determination of the specific heat
amplitude ratio is especially worthwhile for two reasons. First, we would
like to know whether the universality class reduction also holds for this
amplitude ratio. As this property is also obeyed by the susceptibility
amplitude ratio \cite{leite3} one can conjecture that this property might
hold for {\it all} critical amplitude ratios. Secondly, it would be highly
desirable to compare the theoretical value of this amplitude ratio with
experiments carried out in some real magnetic material exhibiting the
Lifshitz critical behavior. This would yield a test for the convenience
(or not) of the theoretical formalism employed in the solution of this
problem. From the phenomenological viewpoint, theoretical and
experimental studies have proved that manganese phosphide ($MnP$) presents
a pure uniaxial Lifshitz point ($m=1$, $d=3$, $N=1$)
\cite{Yo,Be1}. Moreover, the experimental determination of the
specific heat amplitude ratio for $MnP$ was realized thirteen years
ago by means of susceptibility measurements \cite{Be2}. A theoretical
attempt to describe this amplitude ratio was performed using the mean field
approximation \cite{Na-Ab-Fo} by neglecting the contribution of the
fluctuations, but the agreement with the experiment was not achieved.
A proper treatment should include the effect of fluctuations since they play
a nontrivial role in the determination of this particular amplitude ratio.

In this paper the specific heat amplitude ratio near an anisotropic
$m$-axial Lifshitz point is calculated at one-loop level using a
$\lambda \phi^{4}$ field theory combined with new renormalization group and
$\epsilon_{L}$-expansion methods. We shall treat only the especial Ising-like
case $N=1$, since for $N>1$ and below the Lifshitz critical temperature the
appearance of a massless Goldstone mode leads to infrared problems which
require a separate analysis \cite{BLZ}. In the present field-theoretic setting
the long-standing difficulty in this calculation is that it requires the
knowledge of the coupling constant at the fixed point and the specific heat
critical exponent at two-loop level in order to find out this amplitude ratio
beyond the mean field approximation. These objects have recently
been figured out at two-loop level \cite{leite2} permitting, therefore,
the present analysis. It represents the first theoretical report with the
effect of the fluctuations properly included in the specific heat amplitude
ratio for the anisotropic Lifshitz critical behavior. We point out that the
universality class reduction also holds for this amplitude ratio. This is
another step forward towards extending this property to all
other critical amplitude ratios. Furthermore, we estimate it for the uniaxial
case $m=1$ in three-dimensional systems. The result is in very good agreement
with the experimental determination of the specific heat amplitude ratio in
the magnetic compound $MnP$.

\section{Specific heat amplitude ratio}
Since the method of calculation has been described in a previous work
\cite{leite3} and is somewhat standard for ordinary critical systems
\cite{amit}, we will only review briefly the notations and the basic
steps. The bare Lagrangian for the $m$-axial anisotropic Lifshitz critical
behavior reads:
\begin{equation}\label{1}
L = \frac{1}{2}|\bigtriangledown_{m}^{2} \phi_0\,|^{2} +
\frac{1}{2}|\bigtriangledown_{(d-m)} \phi_0\,|^{2} +
\delta_0  \frac{1}{2}|\bigtriangledown_{m} \phi_0\,|^{2}
+ \frac{1}{2} t_{0}\phi_0^{2} + \frac{1}{4!}\lambda_0\phi_0^{4} .
\end{equation}

The Lifshitz critical region is characterized by $\delta_{0}=0$
with the temperature close but not equal to $T_{L}$. We are
going to restrict our analysis using the condition $\delta_{0}=0$
henceafter.

The anisotropic behavior possesses two independent correlation lengths
parallel and perpendicular to the competition axes. They allow two
independent renormalization group flows in momentum space along
directions parallel or perpendicular to the competition axes
\cite{leite1}. Rigorously speaking, we should assign a label to each
renormalized vertex part associated to the flow along spatial
directions parallel or perpendicular to the competing axes, but we
have no need to label the renormalized quantities in this work, since
the fixed point is independent of the flow direction used to
renormalize the theory \cite{leite2}. Consequently, the amplitude
ratios do not depend on what external momenta scale is varied in order
to define the corresponding renormalization group transformation.

The one-loop renormalized Helmholtz free energy
density at the fixed point obtained by using a nonvanishing
quadratic external momenta scale for the $m$-axial Lifshitz critical
behavior is given by \cite{comment}:

\begin{eqnarray}
F(t,M)\;\; =&& \frac {1}{2}t M^{2} + \frac{1}{4!} g^{*} M^{4} +
\frac{1}{4}(t^{2} + g^{*} t M^{2} + \frac{1}{4} (g^{*} M^{2})^{2})
I_{SP} \nonumber\\
&& + \int d^{d-m}{q}d^{m}k
[\bigl(ln(1 + \frac{t + \frac{1}{2} g^{*} M^{2}}{q^{2} +
(k^{2})^{2}}) - \frac{g^{*} M^{2}}{2(q^{2} + k^{4})}\bigr]\;\;,
\end{eqnarray}
where in the above equation $t, M$ ($t_{0} = Z_{\phi^{2}}^{-1}t,
\phi = Z_{\phi}^{-\frac{1}{2}}M$) are the renormalized (bare) reduced
temperature and order parameter, respectively, $Z_{\phi^{2}},
Z_{\phi}$ are normalization functions, $g^{*}$ is the renormalized
coupling constant at the fixed point, $\vec{q}$ is a
$(d-m)$-dimensional wave vector perpendicular to the competing axes,
whereas $\vec{k}$ is a $m$-dimensional wave vector whose components
are parallel to the competition axes. The integral $I_{SP}$ is defined
by:
\begin{equation}
I_{SP} =  \int \frac{d^{d-m}q d^{m}k}{[\bigl(((k+K')^{2}\bigr)^2 +
(q + P)^{2}] \left( (k^{2})^2 + q^{2}  \right)}\;\;\;.
\end{equation}

We choose to renormalize the theory using normalization conditions
where the external momenta scale are zero ($K'=0$) along the competing
axes whereas their nonvanishing components are perpendicular to the
competition axes. A convenient symmetry point for this integral is
chosen at $P^{2}= \kappa_{1}^{2}=1$. As the dimensionful coupling
constant is related to that dimensionless by
$g^{*} = (P^{2})^{-\frac{\epsilon_{L}}{2}} u^{*}$, where
$\epsilon_{L} = 4 + \frac{m}{2} - d$, this symmetry point
transforms the dimensionful into the dimensionless coupling
constant. The typical geometric angular factor for each loop integral
characterizing the $m$-axial Lifshitz behavior is
$\frac{1}{4}S_{d-m}S_{m} \Gamma(2 - \frac{m}{4})
\Gamma(\frac{m}{4})$. These factors should be extracted whenever a
loop integral is performed and absorbed in a redefinition of the
coupling constant \cite{leite1,leite2,AL1}. At the symmetry point the
integral calculated at nonzero external momenta is given by
$I_{SP} = \frac{1}{\epsilon_{L}}(1 + [i_{2}]_{m}\epsilon_{L})$, where
$[i_{2}]_{m} = 1 + \frac{1}{2}(\psi(1) - \psi(2-\frac{m}{4}))$ and
$\psi(z) = \frac{d ln \Gamma(z)}{dz}$ \cite{AL1}.

Since the
vertex part $\Gamma_{R}^{(0,2)}$ is additively renormalized, the
singular part of the specific heat scales with the temperature in the
form \cite{leite1,leite2}:
\begin{equation}
C= -A|t|^{-\alpha_{L}} = - \frac{\nu_{L2}}{\alpha_{L}} B(u^{*})
- \Gamma_{R}^{(0,2)} ,
\end{equation}
where $B(u^{*})$ is the inhomogeneous term of the renormalization
group equation for $\Gamma_{R}^{(0,2)}$ and
\begin{equation}
\Gamma_{R}^{(0,2)} = \frac{\partial^{2} F(t,M)}{\partial t^{2}},
\end{equation}
is the vertex part which is related to the specific heat only at zero
external momentum insertion.

Above the Lifshitz temperature, $M=0$ and using Eqs. (2) and (5)
we obtain:
\begin{equation}
\Gamma_{R}^{(0,2)} = -\frac{1}{2} \int \frac{d^{d-m}q d^{m}k}{(q^{2}
+ (k^{2})^{2} + t)^{2}} + \frac{1}{2}I_{SP}.
\end{equation}
When $T<T_{L}$, we replace the value of $M$ at the coexistence curve,
where $u^{*} M^{2} = -6t$, to find
\begin{equation}
\Gamma_{R}^{(0,2)} = -\frac{3}{u^{*}}
- 2 (\int \frac{d^{d-m}q d^{m}k}{(q^{2}
+ (k^{2})^{2} + 2|t|)^{2}} - I_{SP}).
\end{equation}
The calculation of the one-loop remaining integrals are
straightforward. The integration over the quadratic loop momenta is
trivial, and when the remaining quartic loop integral integral is
performed we find:
\begin{equation}
\int \frac{d^{d-m}q d^{m}k}{(q^{2}
+ (k^{2})^{2} + \tilde{t})^{2}} =
\frac{\tilde{t}^{-\frac{\epsilon_{L}}{2}}}{\epsilon_{L}}(1 -
\frac{\epsilon_{L}}{2}(\psi(2 - \frac{m}{4}) - \psi(1)),
\end{equation}
where we have used the parameter $\tilde{t}$ in order to unify the
language above and below the Lifshitz critical temperature $T_{L}$.

The exponents $\nu_{L2}$, $\alpha_{L}$ and the fixed point (using
normalization conditions) were calculated in
Ref. \cite{leite2} and expressed as:
\begin{eqnarray}
&& \alpha_{L} = \frac{(4 - N)}{2(N + 8)} \epsilon_{L}
- \frac{(N + 2)(N^{2} + 30N + 56)}{4(N + 8)^{3}} \epsilon_{L}^{2} ,\\
&& \nu_{L2} =\frac{1}{2} + \frac{(N + 2)}{4(N + 8)} \epsilon_{L}
+  \frac{1}{8}\frac{(N + 2)(N^{2} + 23N + 60)} {(N + 8)^3} \epsilon_{L}^{2}, \\
&& u^{\ast}=\frac{6}{8 + N}\,\epsilon_L\Biggl\{1 + \epsilon_L
\,\Biggl[ - [i_{2}]_m + \frac{(9N + 42)}{(8 + N)^{2}}\Biggr]\Biggr\}\;\;.
\end{eqnarray}
Now we take the particular value $N=1$. On the other hand,
the additive constant is given by:
\begin{equation}
B(u^{*}) = \kappa_{1} (\frac{\partial
\Gamma^{(0,2)}}{\partial \kappa_{1}})_{|_{P^{2} = \kappa_{1}^{2}
=1}} = - \frac{1}{2} \kappa_{1}^{-\epsilon_{L}}(1 + [i_{2}]_{m}
\epsilon_{L}).
\end{equation}
After taking $\kappa_{1}^{2} = 1$ we obtain up to first order in
$\epsilon_{L}$ the following result:
\begin{equation}
- \frac{\nu_{L2}}{\alpha_{L}} B(u^{*}) =
\frac{3}{2 \epsilon_{L}} + \frac{3 [i_{2}]_{m}}{2} + \frac{1}{4}
+ \frac{87}{58} = \frac{1}{u_{1}^{*}} + 2.
\end{equation}
Using Eqs.(6), (7) and (8) in conjunction with the value of $I_{SP}$
we can easily determine $\Gamma_{R}^{(0,2)}$ above and below
$T_{L}$. After that, use Eqs.(4) and (12) to derive the specific heat
amplitude ratio at one-loop level:
\begin{equation}
\frac{A_{+}}{A_{-}} = \frac{2^{\alpha_{L}}}{4}(1 + \epsilon_{L}),
\end{equation}
where in this formula the specific heat critical index is given at $O(\epsilon_{L}$, i.e., $\alpha_{L}=\frac{\epsilon_{L}}{6}$.

\section{Discussion}
Firstly, the result displays a universal form, since it only depends
on $m$ and $d$. Note that this easily reduces to the usual critical
behavior in the limit $m \rightarrow 0$, therefore confirming the
reduction of the Lifshitz universality class to the Ising-like
universality class. As the susceptibility amplitude ratio presents the
same property at one-loop level, this might be a general feature of
all critical amplitude ratios, including not only thermodynamic
amplitudes but also correlation amplitude ratios as well as
mixed (correlation/thermodynamic) amplitude ratios. Future work will
be devoted to those issues.

On the other hand, this amplitude ratio was determined experimentally
for $MnP$ and found to be $\frac{A_{+}}{A_{-}} = 0.65 \pm
0.05$ \cite{Be2}. Replacing $m=1, d=3$ into the above expression we find
$\frac{A_{+}}{A_{-}} = 0.74$. If we neglect the error bar in the
experiment, the difference among the two results is about $9\%$,
therefore, in very good agreement with the experimental value obtained
for $MnP$. Nevertheless, it is important to mention that this
experimental value of the specific heat amplitude ratio in $MnP$ was
obtained using nonlinear least-square fits to adjust it along with the value of
the critical exponent $\alpha_{L}$. The constraint that the value
of the critical exponents above and below the Lifshitz critical
temperature must be equal ($\alpha_{L}^{+} = \alpha_{L}^{-}$ consistent
with scaling theory) yielded the experimental result
$\alpha_{L} =  0.46 \pm 0.03$, which in turn led to the value
$\frac{A_{+}}{A_{-}} = 0.65 \pm 0.05$ for $MnP$ \cite{Be2}. Moreover,
the different values for the exponents produced by unconstrained fits
near $T_{L}$ were recognized by those authors as a departure from
theoretical predictions, but could also be atributed to
the closeness of the first-order ferromagnetic-helical(fan) transition,
which makes the experimental analysis more difficult. Indeed, the experimental
specific heat exponent is in disagreement with different theoretical
estimates for the uniaxial three-dimensional case whose specific heat
exponent is approximately $0.20$ \cite{Pleim, Diehl, leite1}.

Since the experimental specific heat
amplitude ratio depends upon the specific heat exponent, the best fit which
produced the $\alpha_{L}$ exponent nearly two times larger than that from
theoretical calculations should affect the value of that amplitude ratio
due to error propagation. Of course, a different fit
resulting in another experimental value of $\alpha_{L}$ does modify the
above amplitude ratio due to crossover effects which take place when
the temperature is varied \cite{Be3}. For instance, when the value of
$\alpha_{L}$ is close to $0.270$ in a certain fit, the value of the
corresponding experimental amplitude ratio (fixed by this value of the
specific heat exponent) is given by $\frac{A_{+}}{A_{-}} = 0.435$, but
the system is at $T=132.4K$ \cite{Be3} which is away of the Lifshitz
temperature $T_{L}= (121 \pm 1)K$ for $MnP$ and is affected by the
crossover to the Ising universality class. The funny surprise is that
very close to the Lifshitz critical temperature this error is rather
small when theoretical and experimental estimates for this amplitude
ratio are compared as discussed in the present work, in spite of the
large deviation between the measured specific heat critical exponent
for $MnP$ and calculated values using theoretical tools as discussed
above.

A comparison of theoretical and experimental results of this critical
amplitude ratio with the one associated to the Ising-like universality
class is illuminating. For $N=1,d=3$ the $\epsilon$-expansion
($\epsilon=4-d$) result of this amplitude ratio at one-loop level yielded
$\frac{A_{+}}{A_{-}} = 0.56$ \cite{BLZ}. By neglecting the error bars,
specific heat measures in different materials belonging to the same
universality class resulted in critical amplitude ratios varying from
$0.53$ to $0.58$ \cite{Sa-Le}. On the other hand, series expansion
studies of this ratio produced values in the range $0.62-0.70$
\cite{Ta-Fi}. The $\epsilon$-expansion estimate furnishes better values
than those coming from series expansion even though the $\epsilon$
parameter is not a small number in this case ($\epsilon=1$).

For three-dimensional uniaxial systems $\epsilon_{L} = 1.5$ is even
bigger than in the situation for the Ising-like case analysed
above. This might be the main consequence for a greater difference
between the experimental and theoretical results of this amplitude
ratio when compared to the ordinary critical behavior. In order to
see whether this is a good argument, we suggest the measurement of
this specific heat amplitude ratio in other uniaxial magnetic materials
like $CoNb_{2}O_{6}$ \cite{We}. For Ising-like magnets the deviation of
these specific heat amplitude ratios for different materials is rather
small among themselves as well as when they are compared with results derived
from the $\epsilon$-expansion. Hence, from our $\epsilon_{L}$-expansion
results we would expect a larger deviation of this amplitude ratio for
different materials presenting a uniaxial critical behavior. Nevertheless,
our results are rather encouraging in comparison with experiments in $MnP$,
indicating the reliability of the $\epsilon_{L}$-expansion approach to this
sort of behavior. Since further numerical works have not been pursued for
critical amplitude ratios pertaining to the anisotropic Lifshitz
universality classes, we hope our endeavor can stimulate other
numerical estimates like series expansions or Monte Carlo simulations,
in order to improve the comprehension of as many critical amplitude ratios
as possible for competing systems.

In summary, we have derived the specific heat critical amplitude ratio
above and below the Lifshitz critical temperature for a general
anisotropic behavior $1< d < m-1$. The anisotropic universality class
reduces easily to the Ising-like one in the limit
$m \rightarrow 0$. Needless to say, the result is independent of the
choice of the symmetry point employed: had we started with a symmetry
point characterized by zero external momenta perpendicular to the competing
axes and nonvanishing external momenta components parallel to the competition
axes, we would have arrived at the same result. The uniaxial value of this
ratio is in very good agreement to that measured for $MnP$. Hence, the
analytical approach presented here in order to treat this amplitude ratio
is convenient to describe phenomenological aspects of real physical systems
presenting competing interactions. A thorough study of all critical
amplitude ratios as well as a nonperturbative proof of their universal
character based solely on renormalization group arguments for both
anisotropic and isotropic cases will be an interesting topic for
future investigation.

The author thanks C. C. Becerra for useful discussions and financial
support from FAPESP, grant number 00/06572-6.

\end{document}